# Analytical Solutions for Beams Passing Apertures with Sharp Boundaries

Eitam Luz[1], Er'el Granot[2], and Boris A. Malomed[1]

[1]Department of Physical Electronics, School of Electrical Engineering, Faculty of Engineering, Tel Aviv University, Tel Aviv 69978, Israel
[2]Department of Electrical and Electronic Engineering, Ariel University of Samaria, Ariel, Israel,

*Corresponding author: eitamluz@tau.ac.il.

An approximation is elaborated for the paraxial propagation of diffracted beams, with both one- and two-dimensional cross sections, which are released from apertures with sharp boundaries. The approximation applies to any beam under the condition that the thickness of its edges is much smaller than any other length scale in the beam's initial profile. The approximation can be easily generalized for any beam whose initial profile has several sharp features. Therefore, this method can be used as a tool to investigate the diffraction of beams on complex obstacles. The analytical results are compared to numerical solutions and experimental findings, which demonstrates high accuracy of the approximation. For an initially uniform field confined by sharp boundaries, this solution becomes exact for any propagation distance and any sharpness of the edges. Thus, it can be used as an efficient tool to represent the beams, produced by series of slits with a complex structure, by a simple but *exact* analytical solution.

## 1. Introduction

Diffraction occurs whenever a light beam is scattered by an obstacle. It is customary to analyze the propagation of light using the scalar diffraction theory, which introduces substantial simplifications compared to the full vectorial theory [1]. The most accurate method available for the scalar treatment is the Rayleigh–Sommerfeld diffraction integral [2]. It yields highly accurate results for both far-field and near-field diffraction [3, 4]. However, in most cases, the integral expressions cannot be calculated in an exact analytical form. Some approximations, such as Fraunhofer's for the far field and Fresnel's for the near field, are used to reduce the complexity of the integral formulas. Moreover, the paraxial approximation (PA) is often used for regions close to the optical axis. However, there are difficulties in applying the PA to the near field [5].

The rapid progresses in the computer technology in the last few decades have led to a revolution in the computational wave optics. Ordinary modern computers can simulate the optical diffraction without the resort to the Fresnel or Fraunhofer approximations [5, 6, 7]. Nevertheless, there is a continuing demand in industry and academia to increase the complexity of beam-propagation models, which requires the use of extremely powerful computation resources, such as digital holography or computer-generated holograms [8, 9]. Therefore, there is a trend to incorporate analytical approximations into numerical calculations, to mitigate the computational complexity [8, 9, 10, 11, 12]. Analytical solutions improve the understanding of the diffraction patterns, and help to develop the intuition necessary for the work with models under the examination. In particular, the analytical methods may be used to establish fundamental limitations of the outcome of the diffraction (see, e.g., Ref. [13]. Moreover, analytical solutions are immune to numerical limitations, such as the sample size and sampling rate.

In Ref. [14], the propagation of coherent beams with initially sharp transverse rectangular boundaries was investigated theoretically and experimentally in the framework of the one-dimensional (1D) PA. That work developed a diffraction counterpart of the Schrödinger dynamics of initially sharply bounded signals [15, 16, 17, 18]. The Schrödinger equation exhibits universal behavior in the short-time regime as a response to sharp boundaries of the input. Similar to the propagation of signals in dispersive media, the paraxial wave equation has a Schrödinger-like structure, therefore it has been shown, both experimentally and theoretically, that the same universal behavior holds in the diffraction domain as well [14].

The universality was used to generate an approximate generic solution, which is valid *in the entire space*. However, it required a division of space into three domains. In a vicinity of the sharp boundary, the evolution may be approximated by a linear function, in the intermediate domain the short-distance approximation (SDA) is valid, while in the far domain the influence of the boundary becomes negligible. It was shown that this generic 1D solution approximates the exact dynamics with good accuracy [14]; however, the solution is neither analytical nor even continuous.

The objective of the present work is to derive a generic *analytical* solution for the diffraction dynamics for beams released from any two-dimensional (2D) orifice with sharp boundaries.

## 2. Sharp boundaries and the paraxial diffraction integral

The basic solution of the 2D paraxial wave equation

$$\frac{\partial^2 a}{\partial x^2} + \frac{\partial^2 a}{\partial y^2} + 2ik\frac{\partial a}{\partial z} = 0, \quad (1)$$

where $k = \omega n/c$ is the wave number and $a(x,z)$ is the slowly varying amplitude (SVA) of the scalar electromagnetic field with carrier frequency $\omega$ ($n$ is the refractive index and $c$ the speed of light in vacuum), is derived using the convolution relation [1]

$$a(x,y,z) = \int_{-\infty}^{+\infty}\int_{-\infty}^{+\infty} a(x',y',0) h_z(x-x', y-y')dx'dy' \quad (2)$$

where $h_z(x,y)$ is the 2D impulse response for the paraxial wave equation

$$h_z(x,y) = \frac{k}{2\pi i z}\exp\left(\frac{ik}{2z}(x^2+y^2)\right). \quad (3)$$

Equation (2) with an additional phase factor, $\exp(ikz)$, is sometimes referred to as the Fresnel (paraxial) diffraction integral [6]. In the 1D limit the equation reduces to

$$\frac{\partial^2 a}{\partial x^2} + 2ik\frac{\partial a}{\partial z} = 0. \quad (4)$$

This case is suitable for a slit geometry, where the medium is stationary and homogeneous, and the beam source is at least partially coherent in accordance with the SVA constrains. The amplitude profile at distance $z$ is given by [1, 19]

$$a(x,z) = \int_{-\infty}^{+\infty} a(x',z=0) h(x-x',z)dx', \quad (5)$$

where the impulse response is

$$h(x,z) = \sqrt{\frac{k}{2\pi i z}}\exp\left(ik\frac{x^2}{2z}\right) \quad (6)$$

An important example is the case when the initial amplitude profile is represented by the Heaviside step function, $u(x)$, multiplied by a constant $a_0$:

$$a(z=0,x) = a_0 u(-x) \quad (7)$$

The substitution of (7) in Eq. (5) yields the solution which is valid for any $z$:

$$\frac{a(z,x)}{a_0} = \frac{1}{2}\text{erfc}\left(\frac{x}{\sqrt{2izk^{-1}}}\right). \quad (8)$$

This solution can be easily generalized for input in which the step function is replaced by its continuous counterpart with a transition length scale $\chi$, namely,

$$u_\chi(z=0,-x) \equiv \frac{1}{2}\text{erfc}\left(\frac{x}{\chi}\right). \quad (9)$$

In this case, the solution is (see Refs. [14, 18] and Appendix A.1)

$$u_\chi(z,-x) = \frac{1}{2}\text{erfc}\left(\frac{x}{\sqrt{2izk^{-1}+\chi^2}}\right) \quad (10)$$

Clearly, Eq. (10) carries over into Eq. (8) in the limit of $\chi \to 0$. Hereafter, $u_\chi(z,x)$ is referred to as the "continuous step function" (CSF).

It was shown in Ref. [14] that, for any initially singular beam profile of the form

$$a(z=0,x) = f(x)u(-x), \quad (11)$$

where $f(x)$ is any arbitrary analytic function, the paraxial diffraction integral yields an exact generic solution:

$$a(z>0,x) = M\left(x, -i\frac{\partial}{\partial \xi}, z\right) f(\xi)\Big|_{\xi=0} \quad (12)$$

where

$$M(x,q,z) \equiv \frac{1}{2}\exp\left(iqx + iq^2\frac{z}{2k}\right)\text{erfc}\left[\frac{x+qzk^{-1}}{\sqrt{-2izk^{-1}}}\right] \quad (13)$$

is the Moshinsky's function [13, 14, 20, 21] and $M\left(x,-i\frac{\partial}{\partial \xi},z\right)$ is an operator, which acts on function $f(\xi)$ at the singular boundary, $\xi = 0$. Therefore, the beam dynamics is determined solely by the envelope shape at the singularity. While, this generic solution does not have a closed-form analytical expression, analytical expressions were derived for the short-distance approximation, $2z \ll kx^2$, in which case

$$a(z>0,x) \approx a(z=0,x) + \frac{1}{x}\sqrt{\frac{iz}{2\pi k}} \exp\left(i\frac{kx^2}{2z}\right) f(0) \quad (14)$$

In the general case of the discontinuous input, with amplitude jumps between arbitrary values, i.e.

$$a(z=0,x) = f_-(x)u(-x) + f_+(x)u(x), \quad (15)$$

the solution is

$$a(z>0,x) \approx a(z=0,x) + \frac{1}{x}\sqrt{\frac{iz}{2\pi k}} \exp\left(i\frac{kx^2}{2z}\right)\Delta a, \quad (16)$$

where $\Delta a \equiv a(0,-0) - a(0,+0) = f_-(0) - f_+(0)$ is the boundary amplitude jump. This solution is the SDA (recall this acronym stands for the short-distance approximation). It is clear that, in this approximation, the beam structure is entirely determined by the amplitude gap. Similarly, if the amplitude is continuous but the singularity is represented by a jump of the first derivative, i.e., $f_+(0) = f_-(0)$ but $\left.\frac{\partial f_-(\xi)}{\partial x}\right|_{\xi=0} \neq \left.\frac{\partial f_+(\xi)}{\partial x}\right|_{\xi=0}$, there another universal approximation is available [15, 18]:

$$a(z,x) \cong a(0,x) - \frac{(izk^{-1})^{3/2}}{x|x|} \sqrt{\frac{1}{2\pi}} \exp\left(i\frac{kx^2}{2z}\right)\left[f'_-(0) - f'_+(0)\right], \quad (17)$$

where the prime stands for $d/dx$. It can be seen from Eqs. (16) and (17) that the discontinuity in the input gives rise to factor $z^{1/2}$ in the universal solution, while, when the slope (derivative) is discontinuous, the factor is a milder one, $z^{3/2}$.

Despite their universality, expressions (16) and (17) are valid only at short distances, which determine a finite spatial layer. To derive an approximate solution for larger distances, the space is separated into three regions. Accordingly, two length scales are defined: the short one,

$$\xi_1 \equiv \left[\left(2k^{-1}z\right)^2 + \chi^4\right]^{1/4}, \quad (18)$$

and the long scale,

$$\xi_2 \equiv 2(k\chi)^{-1} z. \quad (19)$$

With these definitions, a generic solution can be written as

$$a(z>0,x) = \begin{cases} a(0,x) & , \quad |x| > \xi_2 \\ a(0,x) + \frac{\Delta a}{|x|}\sqrt{\frac{iz}{2\pi k}} \exp\left(i\frac{kx^2}{2z}\right) & , \quad \xi_1 < |x| < \xi_2 \\ \frac{\bar{a}}{2}\left[1 - \frac{2}{\sqrt{\pi}} \frac{x}{\sqrt{2iz/k + \chi^2}}\right] & , \quad |x| < \xi_1 \end{cases} \quad (20)$$

where $\bar{a} \equiv (a(0,+0) + a(0,-0))/2$ is the mean value of the discontinuous function at the jump point. In a vicinity of the singularity, i.e., at $|x| < \xi_1$, the linear approximation holds. In the intermediate layer, i.e., at $\xi_1 < |x| < \xi_2$ the SDA is valid, while, at $\xi_2 < |x|$, there is no significant evolution of the beam.

Eq.(12) and (20) were generalized to the general case of the propagation of singular beams in *any* linear medium and especially to third order dispersion medium in Ref. [22].

It has been found [14, 18] that this approximation (20) shows good agreement with numerical simulations and experimental findings. However, despite its relative simplicity, a drawback of this approximation is the fact that it is not analytical. In fact, in most cases it is not even continuous.

Below we develop an analytical approximation for the propagation of beams released from apertures with sharp boundaries, which will be valid for the entire space.

### 3. The Analytical Approximation

To derive an analytic approximation for the propagation of a discontinuous signal, we can take advantage of three facts:
1) There is the exact analytical solution (8) for the propagation of the initial step-function profile (7).
2) The higher is the degree of the singularity, the milder is the dependence on the propagation distance at the initial stage. That is, the propagation of the electromagnetic field initiated by the discontinuous input is determined by factor $\sim z^{1/2}$, see Eq. (16), while the continuous input with a discontinuity in the first derivative gives rise only to $z^{3/2}$ in Eq. (17).
3) In relevant physical settings, boundaries of the initial beam's profile may be very sharp, but they are never truly singular. Moreover, whenever there is a discontinuity in the initial profile, it is always possible to present the profile as a superposition of a continuous function (albeit, not necessarily given in an analytical form) and a simple step function. In particular, Eq. (11) can be rewritten as

$$a(z=0,x) = \left[f(x) - f(0)\right]u(-x) + f(0)u(-x) \quad (21)$$

The second term here represents the simple discontinuity, which gives rise to the exact analytical solution (8), while the first term is continuous, hence it undergoes much milder evolution, which, for the short-distance approximation, may be neglected.

In realistic physical settings, the beam's boundaries, no matter how sharp they are, always have a finite thickness. Therefore, the step functions may be replaced by the CSF with transverse thickness $\chi$, i.e.,

$$a(z=0,x) = [f(x)-f(0)]u_\chi(0,-x)+f(0)u_\chi(0,-x) \quad (22)$$

In this form, both terms in Eq. (22) are continuous, but, since the latter one is characterized by smaller thickness transition, $\chi$, it dominates the ensuing dynamics, while the evolution of the former term may be neglected. Therefore, the beam's profile, after having passed distance $z$, may be approximated by the following *analytical* expression:

$$a(z,x) \approx [f(x)-f(0)]u_\chi(0,-x)+u_\chi(z,-x)f(0)$$
$$= a(0,x)+f(0)[u_\chi(z,-x)-u_\chi(0,-x)], \quad (23)$$

where the CSF factor $u_\chi(z,-x)$ is given by Eq. (10). More generally, when the jump in the input occurs between two arbitrary values, which corresponds to the continuous counterpart of Eq. (15):

$$a(z=0,x) = f_-(x)u_\chi(0,-x)+f_+(x)u_\chi(0,x), \quad (24)$$

where $f_-(x)$ and $f_+(x)$ are arbitrary analytical functions, whose smallest length-scale is considerably larger than $\chi$, which may be arbitrarily small. Then, the same arguments that led to Eq. (22) at $z=0$ produce the following approximation at $z>0$:

$$a(z,x) \approx a(0,x)+\Delta a[u_\chi(z,-x)-u_\chi(0,-x)], \quad (25)$$

where $\Delta a \equiv f_-(0)-f_+(0)$ is the initial jump at $x=0$.

## 4. Beams with multiple sharp boundaries in the 1D and 2D geometries

### 4.1 The 1D setting

When the input contains several jumps, each one of them contributes a term to the solution similar to the last one in Eq. (25). For example, if the initial profile is represented by an arbitrary function $f(x)$, which is bounded in a slit of width $w$, it is written as

$$a(z=0,x) = f(x)\{u_\chi(0,-x-w/2)-u_\chi(0,-x+w/2)\}. \quad (26)$$

After passing distance $z$, the corresponding beam's profile becomes

$$a(z,x) \approx a(0,x)+$$
$$\{f(w/2)[u_\chi(z,-x-w/2)-u_\chi(0,-x-w/2)]- \quad (27)$$
$$f(-w/2)[u_\chi(z,-x+w/2)-u_\chi(0,-x+w/2)]\}.$$

The evolution of the beam, including oscillations and broadening, in described by this *analytical solution*. Figure 1 displays comparison between the analytical solution given by Eq. (27) and the corresponding numerical solution of Eq. (5). In these figures $f(x)=1+0.3[\sin(2\pi x/w)+\sin(4\pi x/w)]$, $w$ is the slit's width, $z_R \equiv kw^2/2$ is Rayleigh length [19], and $I/I_0$ is the ratio between intensity $I=|a(z,x)|^2$ and its initial maximum value $I_0 = \max_x |a(z,x)|^2$. As can be seen from these figures, the approximation is excellent for short distances, but deteriorating, as it might be expected, with increase of propagation distance $z$.

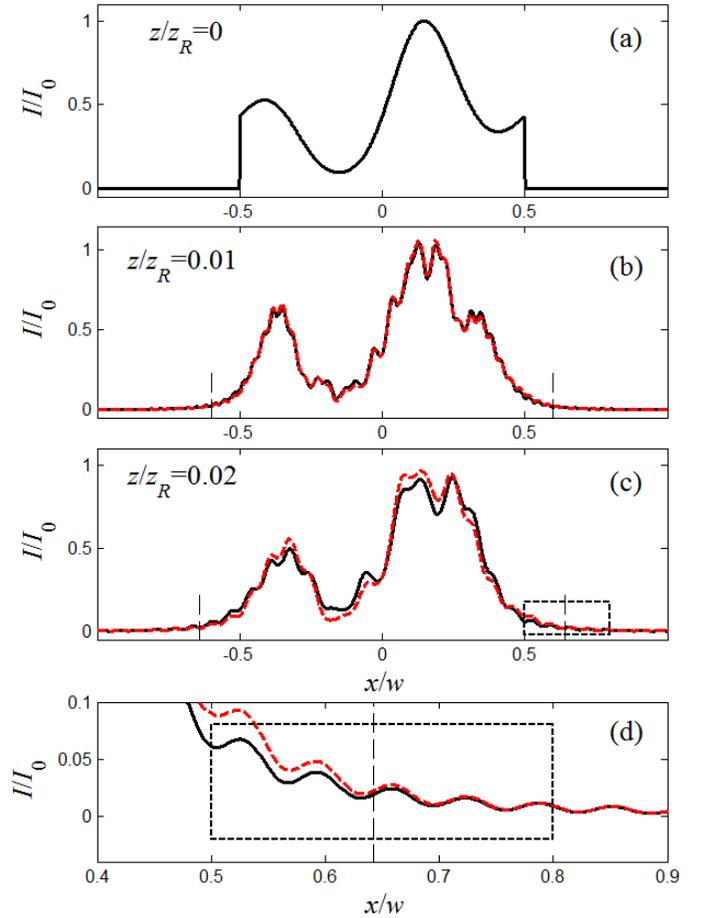

**Figure 1** – (Color online) Comparison between the numerically found intensity (the solid curve) and the analytic approximation given by Eq. (25). The parameters are: $k=10^7$ m$^{-1}$, $w=1$ mm, and $z_R = 4.96$ m. (a) The initial intensity (the solid curve). (b) The numerically found

intensity (the solid curve) and the analytic approximation (the dashed curve). The vertical dashed curves represents $x = -0.5 + \xi_1/w$ and $x = 0.5 + \xi_1/w$ (c) Same as (b) but for a longer distance. (d) Enlargement of the area around the dashed square appeared in panel (c).

In the case where $f(x)$ has the same value $A = f(w/2) = f(-w/2)$ at both edges, Eq. (27) can be simplified to

$$a(z,x) \approx a(0,x) + A\left[R_\chi(z,x,w) - R_\chi(0,x,w)\right] \quad (28)$$

where $R_\chi$ is defined as

$$R_\chi(z,x,w) \equiv u_\chi(z,-x-w/2) - u_\chi(z,-x+w/2). \quad (29)$$

In the same manner, we define

$$R_{\chi_L,\chi_R}(z,x,w) \equiv u_{\chi_L}(z,-x-w/2) - u_{\chi_R}(z,-x+w/2) \quad (30)$$

when the left edge thickness $\chi_L$ is different from the right one, $\chi_R$. This notation allows us to present the propagation of complex beams, shaped by multiple slits, by compact analytical expressions. For example, if $f_n(x)$ is a set of functions that have the same values $A(n)$ at the boundaries, where $n$ determines the location of the $n$th slit, as $x(n) = n\Delta x$, and $w$ is the width of the slits, then the initial SVA can be written as

$$a(z=0,x) = \sum_{n=1}^{N} f_n(x) R_{\chi(n)}(0, x-x(n), w). \quad (31)$$

After passing distance $z$, the solutions is approximately given by

$$a(z,x) \approx a(0,x) + \sum_{n=1}^{N} A(n)\left[R_{\chi(n)}(z,x-x(n),w) - R_{\chi(n)}(0,x-x(n),w)\right]. \quad (32)$$

For a uniform distribution within the boundaries, i.e., $f(x) = A$, the solution becomes exact (similar to the result obtained in the context of the dispersion medium [18]). For example, if input is

$$a(z=0,x) = \sum_{n=1}^{N} A(n) R_{\chi(n)}(0, x-x(n), w), \quad (33)$$

Then, past propagation distance $z$, the solution is *exactly* given by

$$a(z,x) = \sum_{n=1}^{N} A(n) R_{\chi(n)}(z, x-x(n), w). \quad (34)$$

### 4.2. The 2D setting

In the 2D scenario, the impulse response for the free-space paraxial propagation equation is given by Eq.(3), hence the paraxial propagation in free space can be derived as a product of two 1D convolutions. In particular, if the initial state corresponds to a wedge with boundaries defined with the help of CSF's (9),

$$a(z=0,x,y) = u_\chi(z=0,-x) u_\chi(z=0,-y) \quad (35)$$

then the *exact* solution is simply

$$a(z,x,y) = u_\chi(z,-x) u_\chi(z,-y), \quad (36)$$

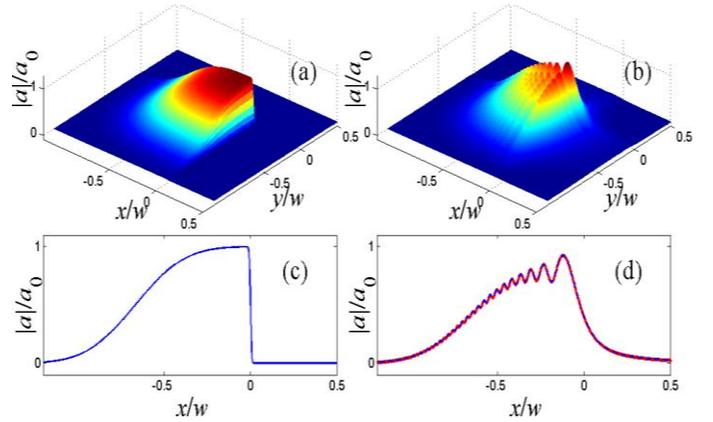

Figure 2 – (Color online) Comparison between the 2D analytical solution given by Eq. (37) and numerical results produced by Eq. (2) for the beam width $w = 1.25$ mm, wavelength $\lambda = 638.2$ nm and propagation distance $z/z_R = 0.006$. The decaying tail is created by a choice of $\chi_L/w = 0.32$, whereas the width of the sharp boundary at the right side is $\chi_R/w = 0.008$. $a_0$ is the maximum value of the input field. (a) The initial beam; (b) the final beam; (c) the cross-section of the input beam at $y = 0$; (d) The numerically found results (solid curve) and the analytic approximation (dashed curve) at the cross-section of the final beam at $y = 0$.

In the same manner, a 2D beam with four corners can be represented by means of the functions defined in Eq. (29):

$$a(z,x,y) = A R_{\chi_L,\chi_R}(z,x,w_x) R_{\chi_L,\chi_R}(z,y,w_y). \quad (37)$$

The analytical solution given by Eq. (37) is displayed in Figure 2 for an asymmetric structure on both dimensions, where on one (say, right) side $\chi_R/w = 0.008$, and on the other side $\chi_L/w = 0.32$. As seen in this figure, this is indeed an exact solution.

In general, a complex set of smooth squares $R_\chi$ can be represented by

$$a(z,x) = \sum_{n,m} A_{n,m} R_\chi(z, x-x(n), w_x(n)) R_\chi(z, y-y(n), w_y(n)), \quad (38)$$

with the initial conditions corresponding to $z=0$. An illustration of the modularity and usability of Eq. (38) is shown in Figure 3, where a complex beam structure may be represented by a simple analytic solution produced by splitting the initial profile into rectangles with sharp boundaries (in this case $\chi = 50\mu m$). Clearly, this method can be applied for any profile, which is initially constructed of multiple rectangles or squares.

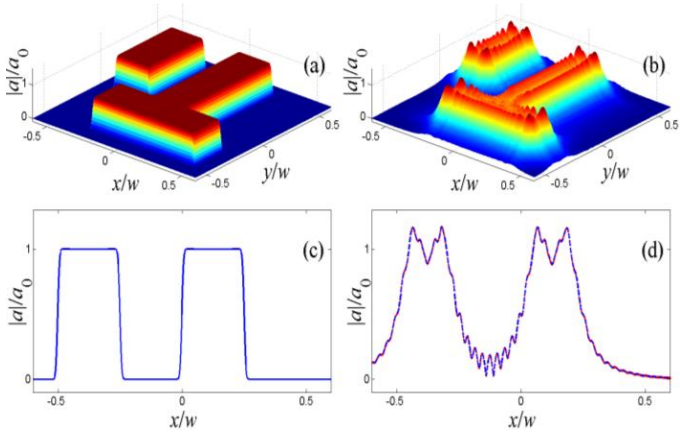

**Figure 3** – (Color online) comparison between the analytical solution given by Eq. (38) and the corresponding numerical solutions of Eq.(2) for wavelength $\lambda = 638.2\,nm$, initial beam width $w = 2.5\,mm$, and propagation distance $z/z_R = 0.024$. The boundary thickness is characterized by $\chi/w = 0.008$, and $a_0$ is the maximum value of the initial field. (a) The initial beam; (b) the final beam; (c) the cross-section of the initial beam at $y/w = 0$; (d) The numerically found results (solid curve) and the analytic approximation (dashed curve) at the cross-section of the final beam at $y/w = 0.4$

## 5. Derivative discontinuity improvement

When the function's derivative does not vanish at the singularity point $(x=0)$, i.e., $f'(x)|_{x=0} \neq 0$, then there is a corresponding discontinuity in the first term of Eq. (21). Nevertheless, as explained above, the effect produced by the latter feature is proportional to $z^{3/2}$, which is negligible for the short distance, compared to the effect of the second (discontinuous) term in Eq. (21), which is proportional to $z^{1/2}$. However, approximation (23) can be improved by rewriting Eq. (21) in a manner that takes account of the derivative, namely,

$$a(z=0,x) = [f(x) - f(0)\exp(\alpha x)]u(-x) + f(0)\exp(\alpha x)u(-x) \quad (39)$$

where $\alpha \equiv f'(x)|_{x=0}/f(0)$. Therefore, the first term in Eq. (39) and its derivatives are continuous at $x = 0$. Similar to Eq. (22) in the non-singular case, i.e., when the boundary thickness is $\chi$, Eq. (39) can be rewritten as

$$a(z=0,x) = [f(x) - f(0)\exp(\alpha x)]u_\chi(0,-x) + f(0)\exp(\alpha x)u_\chi(0,-x), \quad (40)$$

which evolves as

$$a(z>0,x) \approx [f(x) - f(0)\exp(\alpha x)]u_\chi(0,-x) + f(0)U_\chi(z,-x,\alpha), \quad (41)$$

where

$$U_\chi(z,-x,\alpha) \equiv \frac{1}{2}\exp\left[\alpha x - i\alpha^2 \frac{z}{2k}\right]\text{erfc}\left[\frac{x + i\alpha z k^{-1}}{\sqrt{2izk^{-1} + \chi^2}}\right] \quad (42)$$

is the exact solution initiated by input $\exp(\alpha x)u_\chi(0,-x)$ [18]. Further, Eq. (41) can be rewritten as

$$a(z,x) \approx a(0,x) + f(0)\left[U_\chi\left(z,-x,\frac{f'(0)}{f(0)}\right) - U_\chi\left(0,-x,\frac{f'(0)}{f(0)}\right)\right]. \quad (43)$$

More generally, when the initial profile's amplitude jumps between two arbitrary values, as in Eq. (24), a consideration similar to that performed above for Eq. (23) yields (see appendix A.2):

$$a(z,x) \approx a(0,x) + f_-(0)\left[U_\chi(z,-x,\alpha_-) - U_\chi(0,-x,\alpha_-)\right] + \\ -f_+(0)\left[U_\chi(z,-x,\alpha_+) - U_\chi(0,-x,\alpha_+)\right]. \quad (44)$$

In the case of the rectangular slit, i.e.,

$$a(0,x) = [u_\chi(0,-x-w/2) - u_\chi(0,-x+w/2)]f(x), \quad (45)$$

where $f(x)$ is the beam's profile and $w$ the slit's width, the input is, according to Eq. (44),

$$a(z,0) = a(0,x) +$$
$$f(x-w/2)\left[U_\chi(z,-x+w/2,\alpha_{w/2}) - U_\chi(0,-x+w/2,\alpha_{w/2})\right] - \quad (46)$$
$$f(x+w/2)\left[U_\chi(z,-x-w/2,\alpha_{-w/2}) - U_\chi(0,-x-w/2,\alpha_{-w/2})\right]$$

where $\alpha_{-w/2}, \alpha_{w/2}$ are defined by $\alpha_{x_0} \equiv f'(x)\big|_{x=x_0} / f(0)$ for $x_0 = -w/2$ and $w/2$, respectively. Figure **4** demonstrate Eq. (46) for the initial profile $f(x)/f(0) \equiv 4 + 3x/w - \sin(2\pi x/w)$ and an arbitrary slit's width $w$, and compare it to the analytical approximation that neglects the discontinuity in the derivative, i.e., Eq. (27), and to the numerical results.

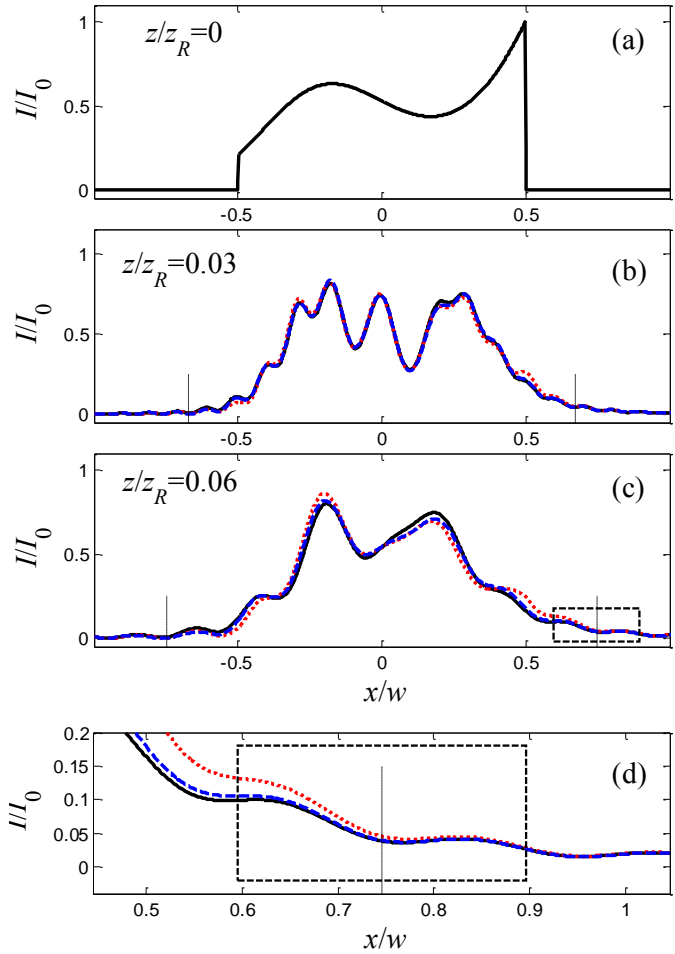

**Figure 4** – (Color online) Comparison between the numerically found intensity (black solid curve) the analytic approximation given by Eq. (27) (red dotted curve) and the Derivative discontinuity improvement given by Eq.(46) (blue dashed curve). The parameters are: $k = 10^7$ m$^{-1}$, $w = 1$mm, and $z_R = 4.96$ m. (a) The initial intensity. (b) Final intensity for $z/z_R = 0.03$ (c) Same as (b) but for a longer distance. The vertical dashed curves in (c) and (d) represents $x = -0.5 + \xi_1/w$ and $x = 0.5 + \xi_1/w$ (d) Enlargement of the area around the dashed square appeared in panel (c).

With a small modification, Eq. (40) can be rewritten for functions that vanish at the transition point, i.e., $f(x=0)=0$, but $\gamma \equiv f'(x)\big|_{x=0} \neq 0$. In this case, we can write for $z=0$
$$a(z=0,x) = \left[f(x) + 1 - \exp(\gamma x)\right]u_\chi(0,-x) - \quad (47)$$
$$\left[1 - \exp(\gamma x)\right]u_\chi(0,-x)$$

And then
$$a(z>0,x) \approx \left[f(x) + 1 - \exp(\gamma x)\right]u_\chi(0,-x) - \quad (48)$$
$$\left[u_\chi(z,-x) - U_\chi(z,-x,\gamma)\right]$$

or
$$a(z>0,x) \approx a(0,x) +$$
$$U_\chi(z,-x,\gamma) - U_\chi(0,-x,\gamma) + u_\chi(0,-x) - u_\chi(z,-x) \quad (49)$$

Figure 5 illustrates this solution for the case of a slit (45) with the profile $f(x)/f(0) = -\sin(2\pi x/w)$, which vanishes at both boundaries of the slits $(x/w = \pm 0.5)$. As can be seen from the figure, it agrees with high accuracy with the numerical solution.

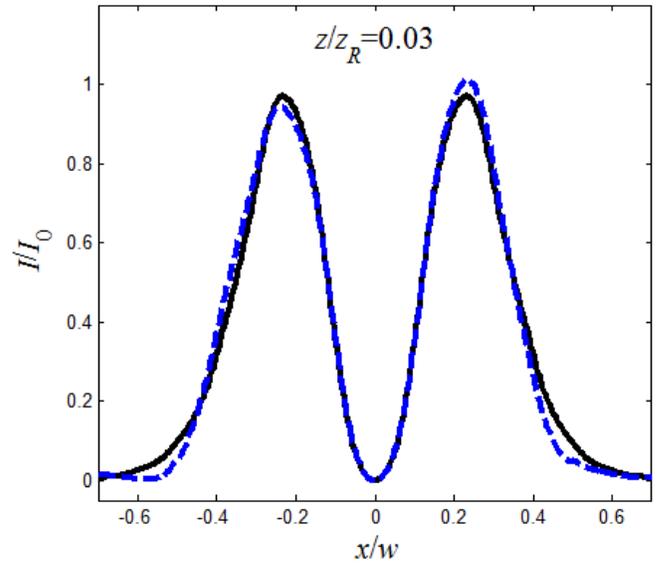

**Figure 5** – (Color online) A beam with sharp features in the derivative profile only. The final numerically generated intensity (the solid curve) is shown along with the analytical solution given by Eq.(46) (the dashed curve). The slit's width is $w = 1$mm, and the Rayleigh length is $z_R = 4.96$ m.

# 6. COMPARISON TO EXPERIMENTAL RESULTS

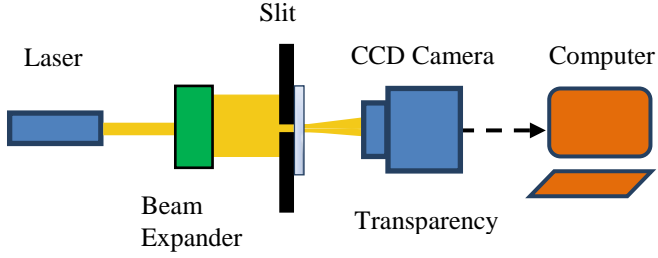

Figure 6: (Color online) A schematic setup.

We have performed experiments, aiming is to compare the observed results and the theoretical approximation corresponding to the SDA and the new approximation given by Eq.(27). The experimental setup (see Figure 6) and parameters are the same as in Ref. [14]. The HeNe laser beam (with wavelength of $\lambda = 632.8\,\mathrm{nm}$) is stretched by a beam expander, and then passed through a 1mm wide slit (1cm long). To control the profile of the beam, the slit is covered by transparencies with different absorption patterns. In this way, the transverse intensity distribution can be manipulated. The transmitted beam is then detected by a CCD camera, and processed with the help of a computer. Figure 7 (symmetric) and

Figure 8 (non-symmetric) show that Eq. (27) indeed produces a good approximation to the experimental results. It is also apparent that this analytical approximation features better agreement with the experimental results than the SDA.

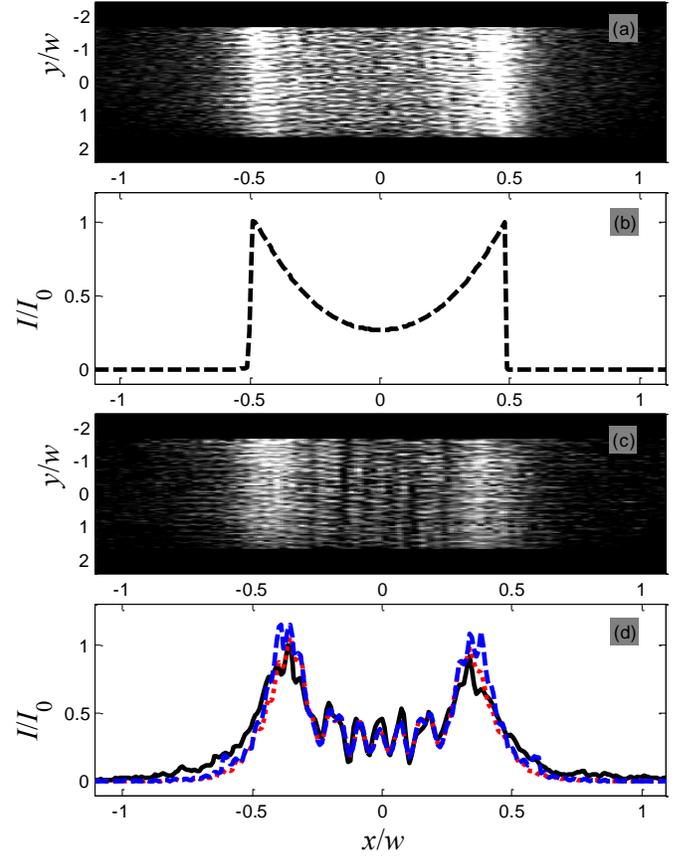

Figure 7 – (Color online) (a) the initial intensity used in the experiment. (b) The averaged (over the y-coordinate) initial intensity. (c) The final intensity produced by the experiment. (d) The comparison between the averaged final experimentally generated intensity (solid curve), the SDA solution (blue dashed curve) and the approximation given by Eq. (27) (the red dotted curve).

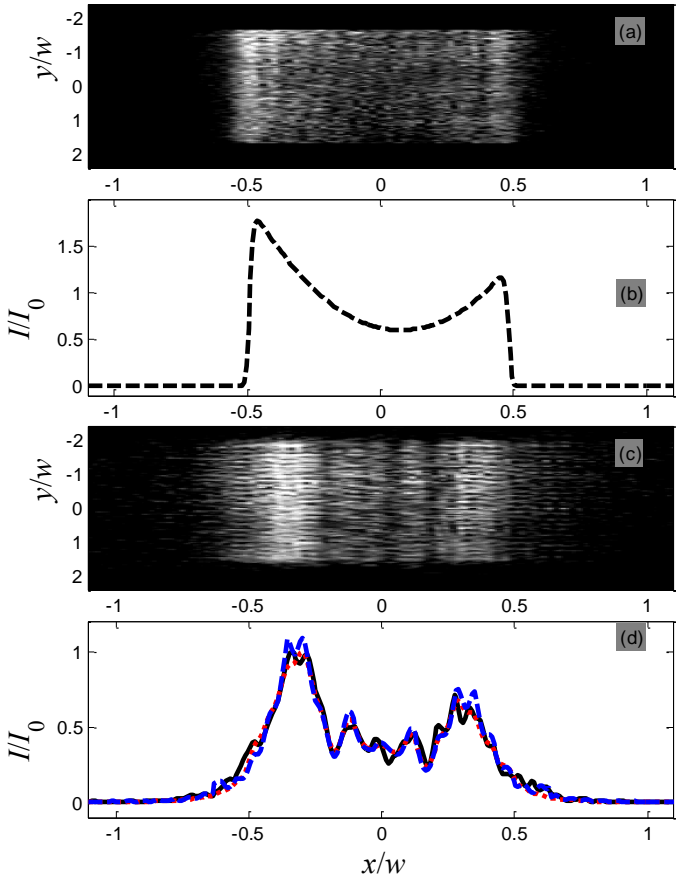

Figure 8 - (Color online) (a) the initial intensity used in the experiment. (b) The averaged (over the y-coordinate) initial intensity. (c) The final intensity produced by the experiment. (d) The comparison between the averaged final experimentally generated intensity (solid curve), the SDA solution (blue dashed curve) and the approximation given by Eq. (27) (the red dotted curve).

## 7. SUMMARY AND CONCLUSIONS

The analytical approximation is elaborated here for generic beams which are launched through apertures with sharp edges. The approximations are valid provided the thickness of the edges is much smaller than any other length scale in the beam's initial profile. However, there is no lower limit for this criterion; the edges can be arbitrarily narrow. The comparison to numerical findings and experimental measurements reveals that this approximation is provided by a superior accuracy over the previous non-analytical approximation. The new solution becomes very accurate for short distances, which are considered the most complicated to solve (analytically and numerically).

The analysis was performed for the 1D and 2D geometries and the results used to represents the propagation of complicated beams, produced by series of slits with a complex structure, by a relatively simple analytical expression. For an initially uniform field within the boundaries, this solution becomes exact for any propagation distance and any sharpness of the edges.

As this approximation is a manifestation of the Schrödinger dynamics of the problem, it may be straight forwardly applied to any system, which is governed by Schrödinger equation(s), such as the propagation in dispersive media, or the propagation of free quantum particles.

## APPENDIX A.1 A DETAILED SOLUTION FOR THE CSF (CONTINUOUS STEP FUNCTION)

The CSF

$$u_\chi(z=0,-x) \equiv \frac{1}{2}\mathrm{erfc}\left(\frac{x}{\chi}\right) \tag{A1}$$

is the convolution of the Heaviside's step function and the Gaussian

$$\frac{1}{\sqrt{\pi\chi^2}}\exp\left(-\frac{x^2}{\chi^2}\right). \tag{A2}$$

As the diffraction of (A1) is the convolution of (A1) with the impulse response (6), this is equivalent to the convolution of the step function with

$$\frac{1}{\sqrt{\pi(\chi^2+2izk^{-1})}}\exp\left(-\frac{x^2}{\chi^2+2izk^{-1}}\right), \tag{A3}$$

which is tantamount to Eq. (A1) when $\chi$ is replaced by $\sqrt{\chi^2+2izk^{-1}}$, i.e.,

$$u_\chi(z,-x) = \frac{1}{2}\mathrm{erfc}\left(\frac{x}{\sqrt{2izk^{-1}+\chi^2}}\right). \tag{A4}$$

## APPENDIX A.2 THE DERIVATIONB OF Eq.(44)

If the initial profile is taken as

$$a(0,x) = f_-(x)u_\chi(0,-x) + f_+(x)u_\chi(0,x), \tag{A5}$$

then adding and subtracting $f_-(0)\exp(\alpha_-x)u_\chi(0,-x)$ and $f_+(0)\exp(\alpha_+x)u_\chi(0,x)$ to Eq. (A5) yields

$$\begin{aligned}a(0,x) =& \left[f_-(x) - f_-(0)\exp(a_-x)\right]u_\chi(0,-x) \\ &+\left[f_+(x) - f_+(0)\exp(a_+x)\right]u_\chi(0,x) \\ &+ f_-(0)\exp(\alpha_-x)u_\chi(0,-x) \\ &+ f_+(0)\exp(\alpha_+x)u_\chi(0,x),\end{aligned} \tag{A6}$$

where $\alpha_- \equiv f'_-(x)\big|_{x=0}/f_-(0)$ and $\alpha_+ \equiv f'_+(x)\big|_{x=0}/f_+(0)$.

Because $u_\chi(0,x) = 1 - u_\chi(0,-x)$, one has

$$a(0,x) = \left[f_-(x) - f_-(0)\exp(\alpha_- x)\right]u_\chi(0,-x)$$
$$+ \left[f_+(x) + f_+(0)\exp(\alpha_+ x)\right]u_\chi(0,-x)$$
$$+ f_-(0)\exp(\alpha_- x)u_\chi(0,-x)$$
$$- f_+(0)\exp(\alpha_+ x)u_\chi(0,-x). \quad \text{(A7)}$$

Due to the same arguments that led to Eq.(23), we may neglect the evolution produced by the first two terms:

$$a(z,x) \approx \left[f_-(x) - f_-(0)\exp(\alpha_- x)\right]u_\chi(0,-x)$$
$$+ \left[f_+(x) + f_+(0)\exp(\alpha_+ x)\right]u_\chi(0,-x)$$
$$+ f_-(0)U_\chi(z,-x,\alpha_-) - f_+(0)U_\chi(z,-x,\alpha_+), \quad \text{(A8)}$$

or, equivalently,

$$a(z,x) = a(0,x) + f_-(0)\left[U_\chi(z,-x,\alpha_-) - U_\chi(0,-x,\alpha_-)\right]$$
$$- f_+(0)\left[U_\chi(z,-x,\alpha_+) - U_\chi(0,-x,\alpha_+)\right]. \quad \text{(A9)}$$

When $\alpha_+ = \alpha_-$ Eq. (23) is evidently reproduced.